\documentclass[pdftex]{article}
\usepackage{icrctc07}

\title{VERITAS: Status and Latest Results}
\shorttitle{VERITAS: Status and Latest Results}
\authors{G. Maier, 
V.A. Acciari, 
R. Amini, 
G. Blaylock, 
S.M. Bradbury, 
J.H. Buckley, 
V. Bugaev, 
Y. Butt, 
K.L. Byrum, 
O. Celik, 
A. Cesarini, 
L. Ciupik, 
Y.C.K. Chow, 
P. Cogan, 
P. Colin, 
W. Cui,  
M.K. Daniel, 
C. Dowdall, 
P. Dowkontt, 
C. Duke, 
T. Ergin, 
A.D. Falcone, 
D.J. Fegan, 
S.J. Fegan, 
J.P. Finley, 
P. Fortin, 
L.F. Fortson, 
D. Gall, 
K. Gibbs, 
G. Gillanders, 
J. Grube, 
R. Guenette, 
G. Gyuk, 
J. Hall, %
D. Hanna, 
E. Hays, 
J. Holder,  
D. Horan, 
S.B. Hughes, 
C.M. Hui, 
T.B. Humensky, 
A. Imran, 
P. Kaaret, 
G.E. Kenny, 
M. Kertzman, 
D. Kieda, 
J. Kildea, 
A. Konopelko, 
H. Krawczynski, 
F. Krennrich, 
M.J. Lang, 
S. LeBohec, 
K. Lee, 
H. Manseri, 
A. McCann, 
M. McCutcheon, 
J. Millis, 
P. Moriarty, 
R. Mukherjee, 
T. Nagai, 
J. Niemiec, 
P.A. Ogden,  
R.A. Ong, 
D. Pandel, 
J.S. Perkins, 
F. Pizlo, 
M. Pohl, 
J. Quinn, 
K. Ragan, 
P.T. Reynolds, 
E. Roache, 
H.J. Rose, 
M. Schroedter, 
G.H. Sembroski, 
A.W. Smith, 
D. Steele, 
S.P. Swordy,  
A. Syson, 
J.A. Toner, 
L. Valcarcel, 
V.V. Vassiliev, 
R. Wagner, 
S.P. Wakely, 
J.E. Ward, 
T.C. Weekes, 
A. Weinstein, 
R.J. White, 
D.A. Williams, 
S.A. Wissel, 
M. Wood  
and
B. Zitzer 
}
\afiliations{see http://veritas.sao.arizona.edu/ for affiliations; email: maierg@physics.mcgill.ca}

\abstract{
VERITAS is an atmospheric Cherenkov telescope array designed to study astrophysical sources of
very-high-energy gamma radiation.
Located in southern Arizona, USA, the array consists of four 12\,m-diameter imaging
Cherenkov telescopes.
All four telescopes have been deployed at the basecamp of the Whipple Observatory
and began full operation in early 2007.
This paper describes the operational status of \mbox{VERITAS}, outlines the initial
performance parameters of the instrument, and presents the latest results that have been obtained.
}

\begin{document}
\maketitle

\begin{figure*}[ht]
\centering
\includegraphics[width=0.85\textwidth]{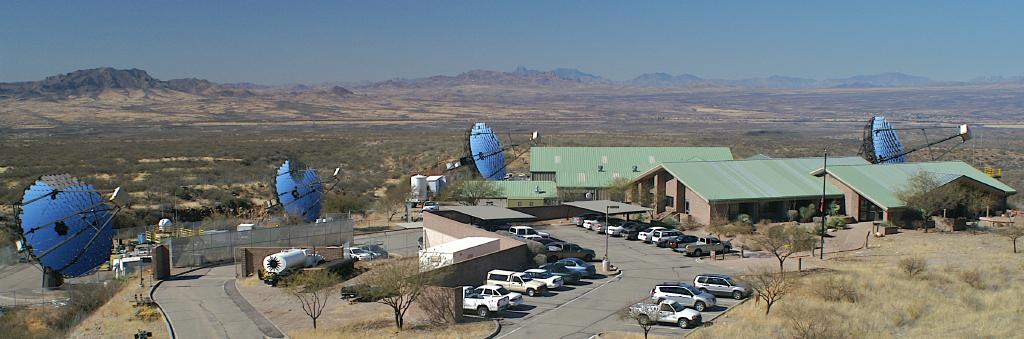}
\caption{\label{fig1}
The \mbox{VERITAS} array of Cherenkov telescopes at the Fred Lawrence Whipple Observatory.
}
\end{figure*}

\vspace{-0.35cm}
\section{VERITAS}
\vspace{-0.30cm}

\mbox{VERITAS} is a ground-based GeV-TeV gamma-ray observatory using an array of four large imaging
Cherenkov telescopes (see Figure \ref{fig1}) \cite{Weekes02}.
It combines a large effective area ($>3\times 10^{4}$ m$^2$) over a wide energy range (100 GeV to 30 TeV)
with good energy (10-20\%) and angular resolution \cite{GM07}.
The high sensitivity of \mbox{VERITAS} allows the detection of sources with a flux of 10\% of the 
Crab Nebula in under 1 hour of observations (5\,$\sigma$ detection).
The angular resolution on an event-by-event basis is better than 0.14$^{\mathrm{o}}$.
Sources of high-energy $\gamma$-rays can currently be located with a precision better 
than 2-3{\tt '} 
(this will improve to below $100''$ with 
the installation of optical pointing monitors on each of the telescopes).
The field of view of the system is 
3.5$^{\mathrm{o}}$ in diameter with an off-axis acceptance above 65\% for offsets smaller
than 1$^{\mathrm{o}}$ from the camera center.
This allows for
the detection of extended sources and is well suited for sky surveys.

\mbox{VERITAS} is located at the basecamp of 
the Fred Lawrence Whipple Observatory in southern Arizona 
(1268 m a.s.l.,~31$^{\mathrm{o}}40'30''$N, $110^{\mathrm{o}}57'07''$W).
The system is operated by an international collaboration of institutions from the U.S.A., Canada, Ireland and the U.K. 
Scientific observations with \mbox{VERITAS} started during the construction phase in spring 2006 with
first two, then three (January 2007), and finally four telescopes (March 2007).
A first light celebration for
\mbox{VERITAS} took place in April 2007.
\mbox{VERITAS} has been fully operational since March 2007 and is now running
an extensive program of scientific observations.

The overall design of each of the four \mbox{VERITAS} telescopes is identical \cite{JH05}.
Each telescope employs a 12\,m-diameter tessellated mirror of Davies-Cotton design with 12\,m focal length
and a total mirror area of 106\,m$^2$.
The reflectivity of the mirrors is measured to be above 90\% at 320 nm \cite{RE07}.
The laser-guided alignment of the 345 mirrors yields a point-spread function of
$\sim0.06^{\mathrm{o}}$ \cite{JT07}.
The focal plane is equipped with a 499-element photomultiplier-tube (PMT) imaging camera \cite{TN07}.
Light cones installed in front of the camera increase the photon collection efficiency and shield
the PMTs from ambient light.
The three-level trigger system of the \mbox{VERITAS} array \cite{AW07,RW07} significantly suppresses background
events due to local muons which are a problem in single-telescope operation.
The typical array trigger rate is 220 Hz with a dead time of $\sim$10\%.
When triggered, the PMT signals in each telescope
are digitized using \mbox{500 MS/s} custom-built FADC electronics \cite{LH07, PC07}.
The actual conditions of the system and at the site are constantly monitored \cite{MH07,MD07b}
and taken into account with a detailed calibration chain \cite{DH07}. 
\mbox{VERITAS} observations can be analyzed by a number of independently developed analysis
packages \cite{MD07,JQ07}.
All aspects of the event reconstruction have been extensively tested with simulations
\cite{GM07}.

\vspace{-0.3cm}
\section{VERITAS observations 2006 - 2007}
\vspace{-0.3cm}

%
%

\begin{figure}
\centering
\includegraphics[width=0.39\textwidth]{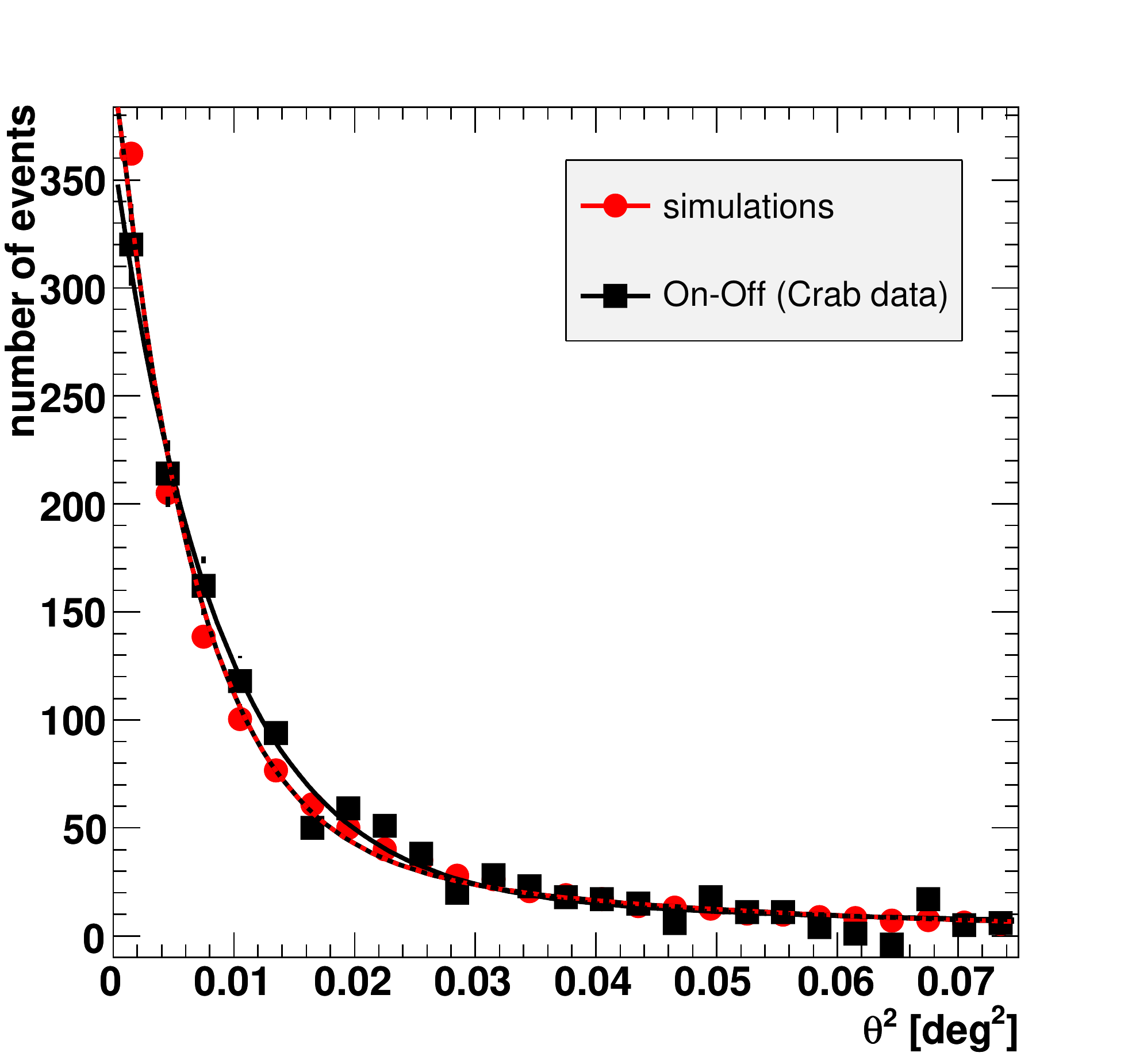}
\caption{\label{fig4} 
$\Theta^2$ distribution for gamma rays recorded from the Crab Nebula in 
comparison with simulations of a point-like source (3-telescope array, see \cite{GM07} for details).
}
\end{figure}

Although the scientific program of \mbox{VERITAS} started in early 2007, several sources
were observed and detected during the construction phase.
The objects studied are mainly part of the \mbox{VERITAS} key science projects, but
comprise as well the observation of $\gamma$-ray bursts \cite{DE07} and research
on cosmic rays \cite{SW07}.
The key science projects comprise supernova remnants and pulsar wind nebulae, dark matter candidates,
blazars, and a survey of three regions of the galactic plane 
with a sensitivity of 5\% of the Crab Nebula flux (beginning in late 
spring 2007).
Some of the early results of the \mbox{VERITAS} science program are listed below;
details and further references can be found in the referred
\mbox{VERITAS} contributions to this conference.

%
%

%

%


\begin{figure}
\centering
%
%
\includegraphics[width=0.36\textwidth]{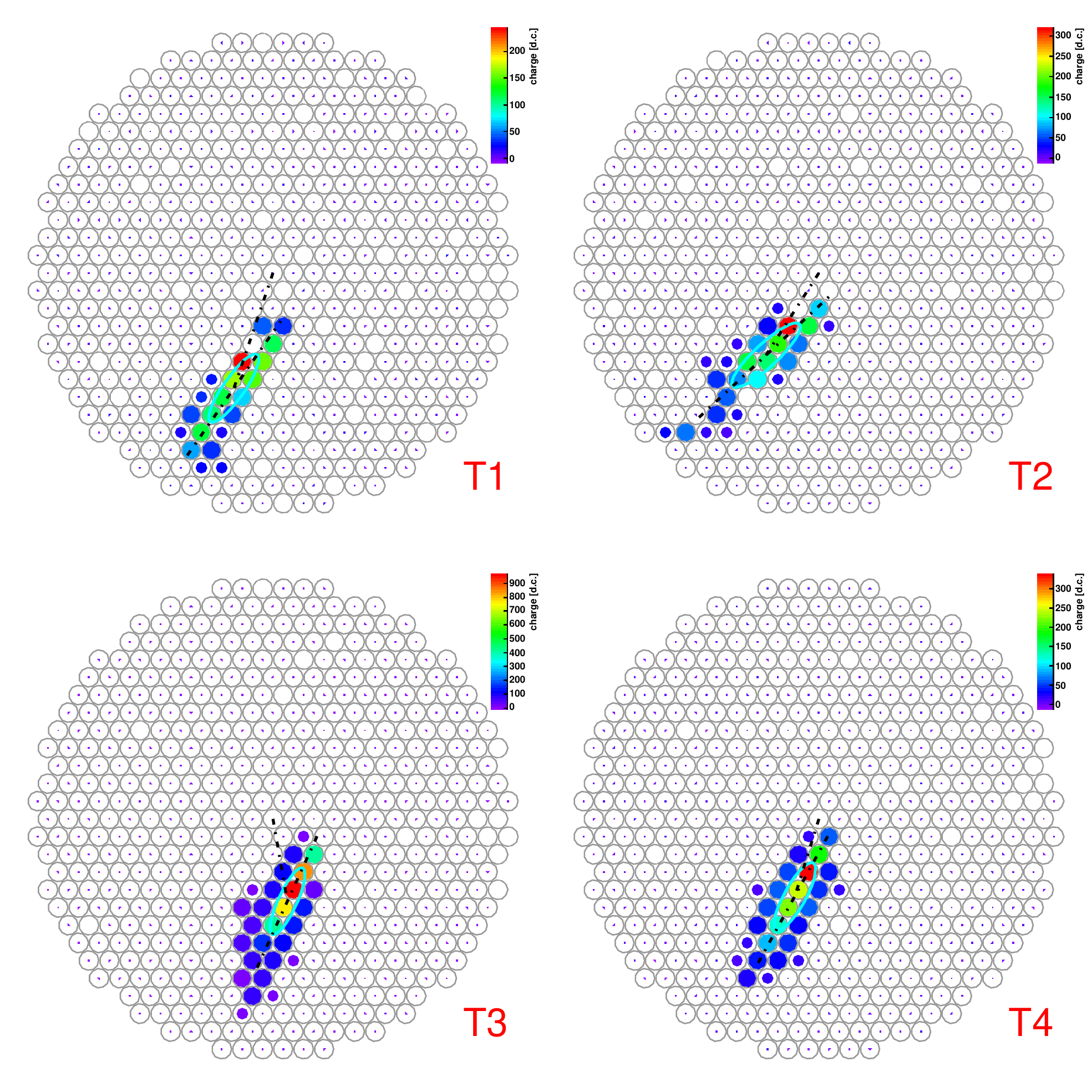}
\caption{\label{fig:4TelEvent}
Camera view of a four-telescope event with \mbox{VERITAS}. 
The color scale indicates the integrated charge per channel.
}
\end{figure}

%
%

\begin{figure}
\centering
\noindent
%
%
\includegraphics[width=0.39\textwidth]{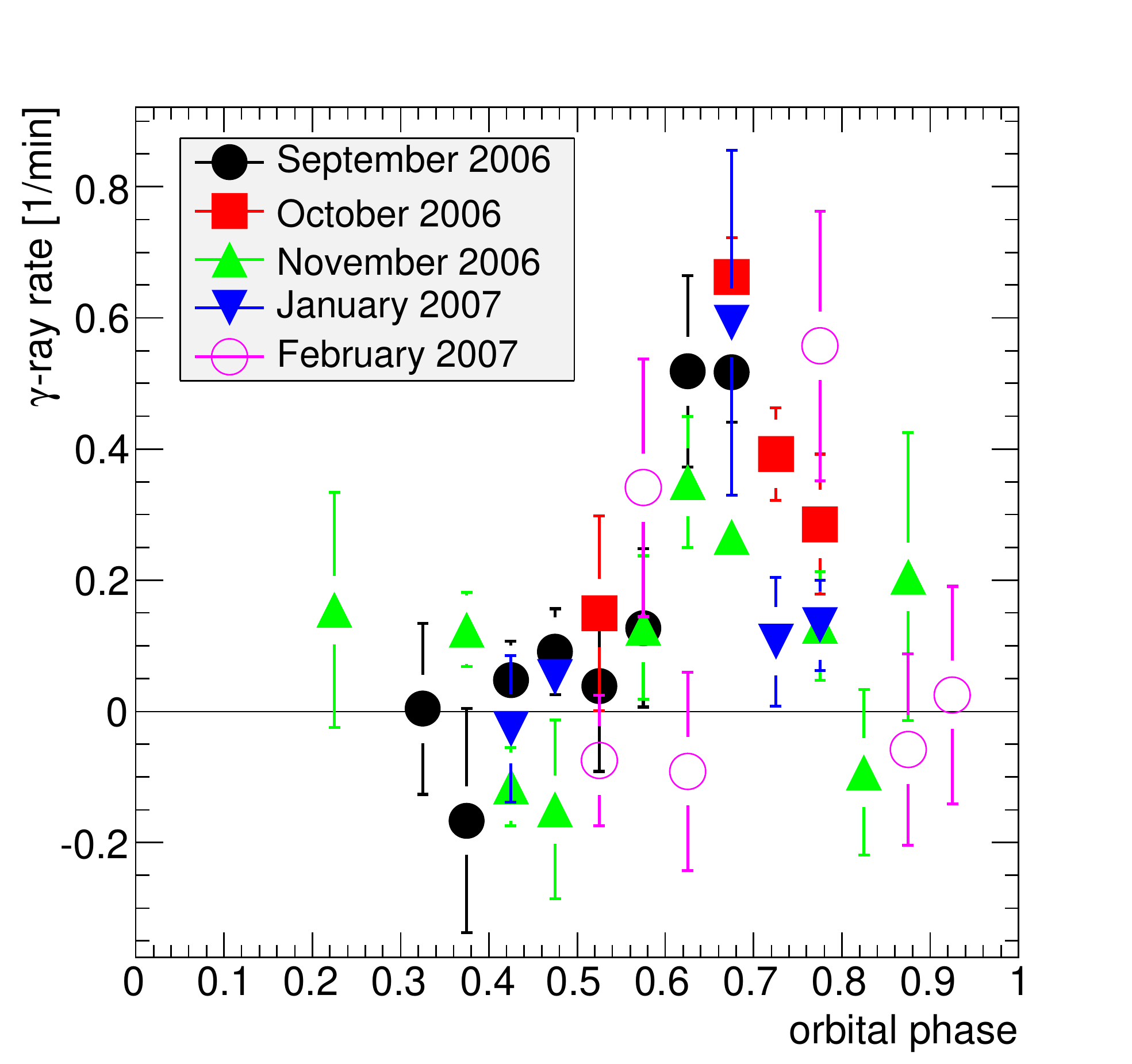}
\caption{\label{fig6}
Rate of excess events versus orbital phase for five different orbits for LS I +61 303.
}
\end{figure}

The Crab Nebula is a strong, steady, point-like source of high-energy $\gamma$-rays.
Observations of this source provide, apart from scientific results, many valuable insights into 
the performance of the instrument.
The Crab Nebula was observed with different array configurations in 2006 and 2007
($\sim$35 h with two telescopes and $\sim$3.5 h with three telescopes).
For the 3-telescope observations, the total detection significance is 51.6\,$\sigma$ with a mean
$\gamma$-ray rate after standard cuts of 7.5 $\gamma$/min (Figure \ref{fig4} and \cite{OC07}).
The measured energy spectrum of the Crab Nebula for the two- and three-telescope data 
is consistent in shape and flux with previous results from other observatories.
Another galactic object observed is the binary system \mbox{LS I +61 303}, a
known emitter of high-energy $\gamma$-rays, detected by MAGIC \cite{Albert06}.
\mbox{LS I +61 303} was detected by \mbox{VERITAS} during several orbital cycles; the total significance
for the whole data set is 8.8\,$\sigma$ \cite{GM07b}.
The observations have been accompanied by extensive X-ray monitoring with Swift and RXTE satellites \cite{AS07}.
The measured $\gamma$-ray rate is strongly variable and is modulated at an
orbital period of 26.5 days,
showing the strong dependence of particle acceleration and/or propagation
on the relative position of the two objects in the system.
The maximum flux appears in each orbital cycle approximately at apastron and
corresponds to about 10\% of the flux of the Crab Nebula
(Figure \ref{fig6}).

Several supernova remnants and pulsar wind nebulae were observed \cite{BH07,AK07};
a highlight is the study of the source SNR IC 443.
A number of potential TeV sources identified by the MILAGRO observatory \cite{Abdo2007}
were coincidentally
in the field of view during several hours of observation with \mbox{VERITAS};
new upper flux limits for these are presented in \cite{DK07}.


\begin{figure}
\centering
%
\includegraphics[width=0.39\textwidth]{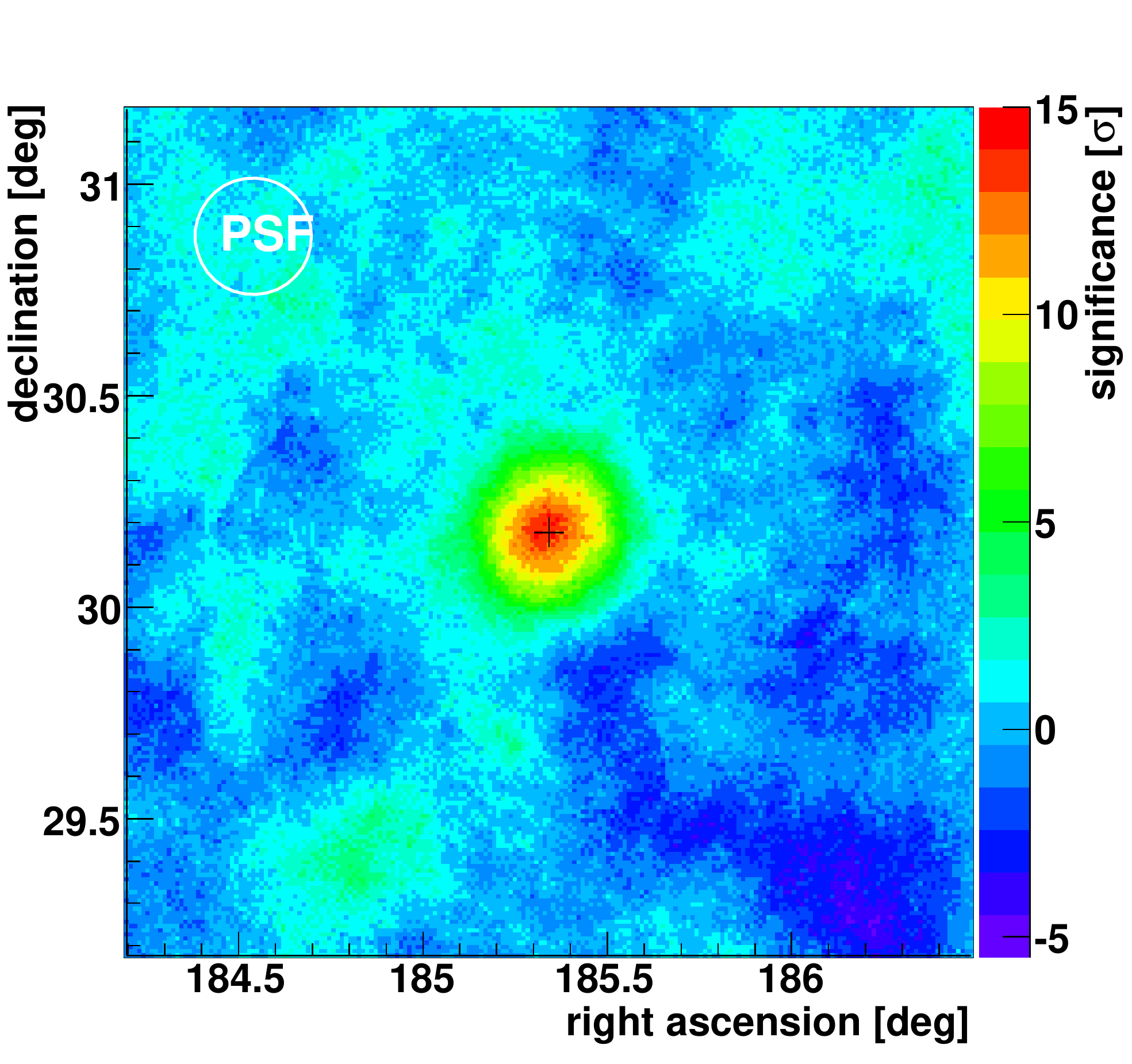}
\caption{\label{fig8}
Significance map of the region around 1ES 1218 +304.
}
\end{figure}

\begin{figure}
\centering
%
%
%
\includegraphics[width=0.39\textwidth]{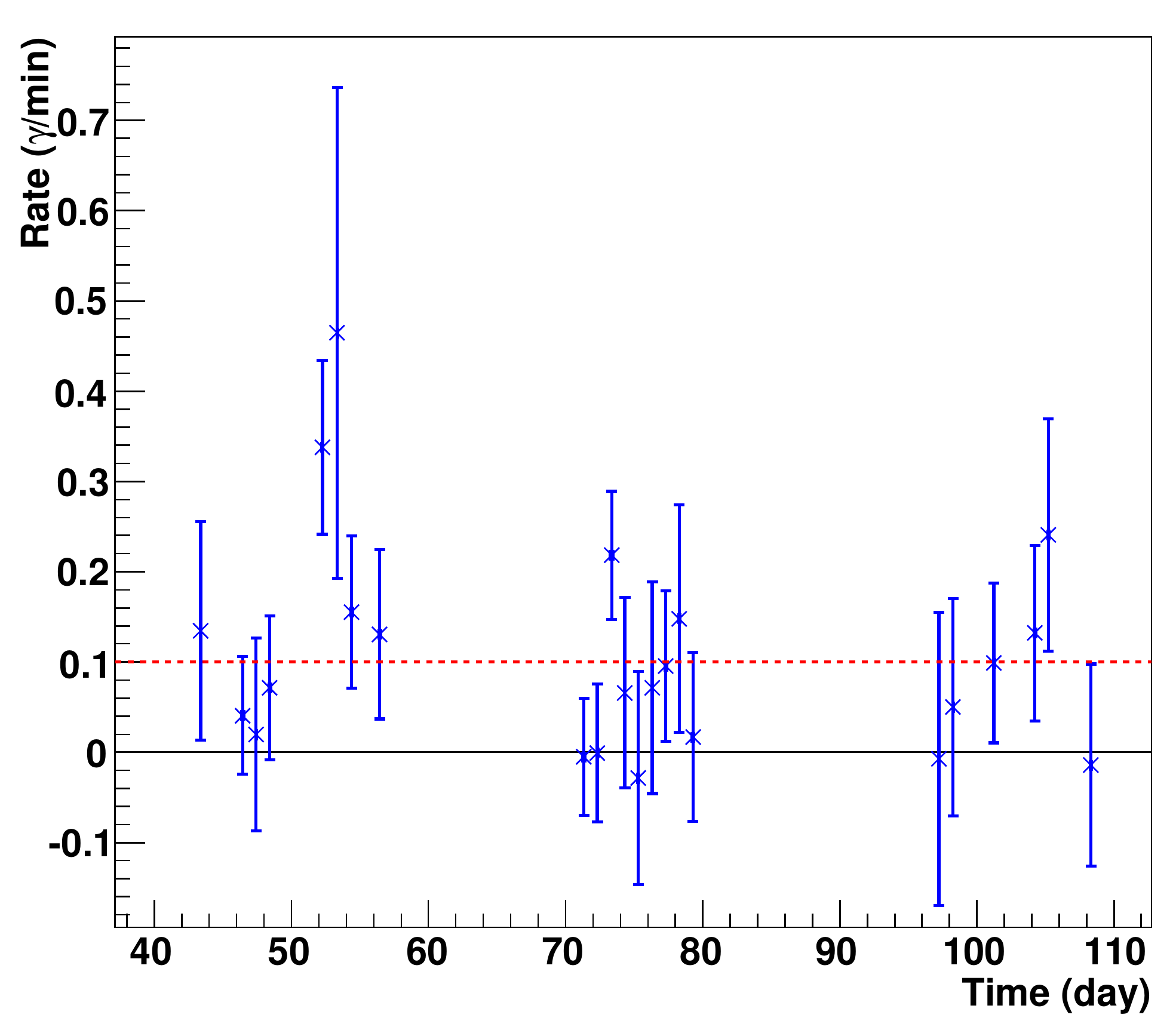}
\caption{\label{fig7}
Light curve of very-high-energy $\gamma$-rays from the direction of M87.
}
\end{figure}

Various extra-galactic sources were observed with \mbox{VERITAS}.
While new upper limits could be set on a number of AGN \cite{HK07, PC07b},
several were clearly detected (Markarian 421 and Markarian 501 \cite{SF07},
\mbox{1ES 1218 +304} \cite{PF07}, and M87 \cite{Pi07}).
\mbox{1ES 1218 +304} is, with a redshift of z=0.182, one of the most distant
blazars observed in high-energy $\gamma$-rays \cite{Albert06b}.
It was detected by \mbox{VERITAS} during $\sim$17 h of observation with a total
significance of 10\,$\sigma$ and a typical $\gamma$-ray rate of 0.3 $\gamma$/min 
(Figure \ref{fig8}).
M87, on the other hand, is the only active galactic nucleus seen in high energy $\gamma$-rays which is
not of the blazar type.
The H.E.S.S. Collaboration recently reported rapid variability and an unexpectedly hard TeV energy spectrum 
for M87 \cite{Aharonian2006b}.
The source has been detected with \mbox{VERITAS} and Figure \ref{fig7} shows the light curve for
early 2007.
The observation of blazars with \mbox{VERITAS} is closely connected to the AGN monitoring program
of the Whipple telescope \cite{DS07}.

\vspace{-0.3cm}
\section{Summary}
\vspace{-0.3cm}

The construction of the \mbox{VERITAS} array of imaging Cherenkov telescopes is now completed;
the system has achieved excellent performance,
with a measured sensitivity for the 3-telescope array corresponding to 
10\% of the Crab Nebula flux in 1.2 h and a Monte Carlo derived sensitivity for
four telescopes of under 1 h for the same flux level.
The various contributions of the \mbox{VERITAS} collaboration to this conference
demonstrate the great potential of the system
for the discovery of new $\gamma$-ray sources and for detailed morphological and spectral studies.

\vspace{-0.35cm}
\subsubsection*{Acknowledgments}
\vspace{-0.30cm}
\begin{small}
This research is supported by grants from the U.S. Department of Energy,
the U.S. National Science Foundation,
and the Smithsonian Institution, by NSERC in Canada, by PPARC in the UK and
by Science Foundation Ireland.
\end{small}

\end{document}